\documentclass[preprint,groupedaddress,floatfix,%
nofootinbib]{revtex4}

\pdfoutput=1 

\usepackage{dcolumn}
\usepackage{graphicx}
\usepackage{epsfig}
\usepackage{hyperref}
\usepackage{graphicx}
\usepackage{dcolumn}
\usepackage{amssymb,eucal}
\usepackage{longtable}
\usepackage{amsmath,amssymb,bm}
\usepackage{mathrsfs}

\newcommand{\mathd}{\mathrm{d}}
\newcommand{\mathi}{\mathrm{i}}
\newcommand{\mathe}{\mathrm{e}}

\newcommand\tr{{\rm tr }}

\newcommand{\nn}{\nonumber}

\def\tr{\mathop{\rm tr}\nolimits}

\begin{document}
\title{Hagedorn transition and topological entanglement entropy}
\author{Fen Zuo\footnote{Email: \textsf{zuofen@hust.edu.cn}}}
\affiliation{School of Physics, Huazhong University of Science and Technology, Wuhan 430074, China}
\author{Yi-Hong Gao\footnote{Email: \textsf{gaoyh@itp.ac.cn}}}
\affiliation{State Key Laboratory of Theoretical Physics, Institute of Theoretical Physics,
Chinese Academy of Sciences, P.O. Box 2735, Beijing 100190, China}

\begin{abstract}

Induced by the Hagedorn instability, weakly-coupled $U(N)$ gauge theories on a compact manifold exhibit a confinement/deconfinement phase transition in the large-$N$ limit. Recently we discover that the thermal entropy of a free theory on $\mathbb{S}^3$ gets reduced by a universal constant term, $-N^2/4$, compared to that from completely deconfined colored states. This entropy deficit is due to the persistence of Gauss's law, and actually independent of the shape of the manifold. In this paper we show that this universal term can be identified as the topological entangle entropy both in the corresponding $4+1 D$ bulk theory and the dimensionally reduced theory. First, entanglement entropy in the bulk theory contains the so-called ``particle'' contribution on the entangling surface, which naturally gives rise to an area-law term. The topological term results from the Gauss's constraint of these surface states. Secondly, the high-temperature limit also defines a dimensionally reduced theory. We calculate the geometric entropy in the reduced theory explicitly, and find that it is given by the same constant term after subtracting the leading term of ${\mathcal O}(\beta^{-1})$. The two procedures are then applied to the confining phase, by extending the temperature to the complex plane. Generalizing the recently proposed $2D$ modular description to an arbitrary matter content, we show the leading local term is missing and no topological term could be definitely isolated. For the special case of ${\mathcal N}=4$ super Yang-Mills theory, the results obtained here are compared with that at strong coupling from the holographic derivation.
\end{abstract}
 \maketitle

\section{Motivation}
 In recent years intensive studies on entanglement entropy~(EE) show that gauge theories possess nontrivial entanglement pattern~\cite{ZCZW:2015}. Practically, different patterns of entanglement could then be employed to classify different phases of gauge theory. In the case of discrete gauge groups, topological order~(TO) characterizes different phases of the gauge theories~\cite{Wen2004}
 ~\footnote[1]{A pedagogic introduction can be found in Part III of ~\cite{ZCZW:2015}. Full classifications, including many recent  progresses, of $2+1D$ bosonic/fermionic TO's are given in~\cite{Wen:2015B,Wen:2015F}.}. More explicitly, the deconfined phase of discrete gauge theory exhibits nontrivial TO. TO is captured by a set of topological data, and one of them is topological entanglement entropy~(TEE), the topological term extracted from the large-size expansion of EE~\cite{Kitaev:2005dm,Levin2006}. A non-vanishing TEE is intimately related to the global gauge-invariant constraint in the deconfined phase, according to the string-net condensation mechanism~\cite{Levin:2004mi}. Generalization of TO to gauge theory with a continuous gauge group seems extremely urgent, but tunrs out not so straightforward. In strong-coupled gauge theories with a gravity dual~\cite{Maldacena:1997re,Gubser:1998bc,Witten:1998qj}, the entanglement entropy can be conveniently calculated with the Ryu-Takayanaki formula~\cite{Ryu:2006bv,Ryu:2006ef}, and attains manifestly the area-law term. The holographic EE behaves quite differently at different subregion scales in a confining background~\cite{Nishioka:2006gr,Klebanov:2007ws}. When properly extended to finite temperature, it also changes discontinuously across the deconfining temperature~\cite{Fujita:2008zv}. However, the holographic derivation gives a vanishing result for the topological term~\cite{Pakman:2008ui}. It is suggested including of $1/N$ corrections~\cite{Hayden:2011ag}, or bulk gravitational anomaly~\cite{Parnachev:2015nca} could induce a nonzero result.

String-net condensation provides a natural physical mechanism for TO in gauge theories~\cite{Levin:2004mi}. Intuitively, there are two necessary ingredients for string-net condensation: deconfined gauge degrees of freedom and global gauge-invariant constraint~\cite{Levin:2005vf}. A simple system which satisfies these two conditions is the high-temperature deconfined phase of a weakly-coupled $U(N)$ gauge theory on a compact manifold~~\cite{Sundborg:1999ue,Aharony:2003sx}. For simplicity, we consider the free theory on $\mathbb{S}^3$. The question now becomes, how is the nontrivial topological property in such a thermal phase related to a nontrivial TEE defined at zero temperature? There are two different ways to think about this. First, consider the entanglement entropy of a $4+1~D$ bulk gauge theory with the sphere $\mathbb{S}^3$ as the entangling surface. The partition induces non gauge-invariant degrees of freedom on the surface. Previous studies show that it is just the global Gauss's law on the entangling surface that gives rise to the topological term~\cite{Levin2006,Pretko:2015}. Secondly, the high-temperature limit of the thermal phase defines naturally a dimensionally reduced theory on $\mathbb{S}^3$. TEE in this theory will then be given by the logarithm of the vacuum partition function, which is also well studied in the case of $2+1 D$ Chern-Simons gauge theories~\cite{Dong:2008ft,Fujita:2008zv}. In this paper we will try to make the relation clear from both points of view, and show that all of them are consistent. In comparison, we also perform the calculation in the confining phase.

The main results in the deconfined phase will be derived in the next section. In section \ref{sec:ConP} we apply the same procedure to the confining phase. Some discussions are given in the last section.



\section{Deconfined phase}

\subsection{Thermal entropy and Gauss's law}

We start with the deconfined phase of $U(N)$ gauge theories on $\mathbb{S}^3$, which has been studied intensively in the literature~\cite{Sundborg:1999ue,Aharony:2003sx}. To derive the full partition function, first one construct the single-particle partition functions from the spectrum of individual field:
\begin{equation}
z(q)\equiv \sum _{i} \mathe ^{-\beta E_i/R}=\sum _{i} q^{E_i}, ~~~~q\equiv \mathe^ {-\beta/R},
\end{equation}
with $\beta$ inverse of the temperature $T$. The explicit expressions for vectors, conformally-coupled scalars and chiral fermions have been derived in various ways~\cite{Sundborg:1999ue,Polyakov:2001af, Aharony:2003sx}
\begin{equation}
z_v(q)=\frac{6q^2-2q^3}{(1-q)^3},\quad z_s(q)=\frac{q+q^2}{(1-q)^3}, \quad z_f(q)=\frac{4 q^{3/2}}{(1-q)^3}.\label{eq.z-single}
\end{equation}
 The derivation is nicely summarized in the appendix of~\cite{Basar:2014jua}. The Gauss's constraint on $\mathbb{S}^3$ can be conveniently imposed through a matrix integral
\begin{equation}
Z(\beta)=\int [\mathd U] \exp\left\{\sum_{k=1}^\infty \frac{z_k}{k}~\tr (U^k)\tr ((U^\dagger)^k)\right\}. \label{eq.Z}
\end{equation}
where
\begin{equation}
z_k\equiv z_v(q^k)+n_sz_s(q^k)-(-1)^k n_fz_f(q^k),\label{eq.zk}
\end{equation}
with $n_s$ the number of scalars and $n_f$ the number of chiral fermions. The matrix $U$ is the holonomy along the Euclidean time circle, after averaging over the sphere. In the large-$N$ limit, the matrix integral could be solved with the saddle-point method~\cite{Brezin:1977sv}, with different phases characterized by different matrix eigenvalue distribution patterns. Introducing a density function $\rho(\theta)$ for the distribution of the eigenvalues of $U$, the effective action of the saddle-point solution is reduced to
\begin{equation}
S[\rho(\theta)]=-\log Z=\frac{N^2}{2\pi}\sum_{k=1}^\infty |\rho_k|^2 V_k,\label{eq.S}
\end{equation}
where
\begin{eqnarray}
\rho_k&\equiv&\int \mathd \theta~ \rho(\theta) \cos(k\theta)=\tr (U^k)/N \label{eq.rhon}\\
V_k&\equiv &\frac{2\pi}{k} [1-z_k].
\end{eqnarray}
In particular, the first moment $\rho_1$ corresponds to the norm of the Polyakov loop. In the deconfined phase, the density function $\rho(\theta)$ turns out to have nontrivial support only in a fan-type segment in the unit circle, which further shrinks to a single point in the high-temperature limit~\cite{Sundborg:1999ue,Aharony:2003sx}.

The expansion of the effective action is determined by the already-known $z_k$'s, and the eigenvalue distribution moments $\rho_k$. Although $\rho_k$ at the saddle point can be formally solved as an infinite summation, the high-temperature expansion is not easy to obtain. One could approximate the solution by truncating to a few nonzero $z_k$'s, which leads to a typical logarithmic term in the thermal quantities~\cite{Gross:1980he,Wadia:2012fr}\cite{Sundborg:1999ue,Aharony:2003sx}. In a recent paper~\cite{Zuo:2014yga} we successfully obtain the first nontrivial terms in the expansion of $\rho_k$'s and the corresponding terms in the thermal quantities, through a direct infinite summation. In the following we review slightly the derivation.


In the high-temperature limit $\rho(\theta)=\delta(\theta)$, or $\rho_k=1$. So one can separate the effective action into two parts
\begin{eqnarray}
\log Z&=&-N^2\sum_{k=1}^\infty  \frac{|\rho_k|^2}{k} \left[1-z_k\right]= -N^2\sum_{k=1}^\infty  \frac{|\rho_k|}{k} \left[1-z_k\right]\nonumber\\
&=& -N^2\sum_{k=1}^\infty  \frac{1}{k} \left[1-z_k\right]+N^2\sum_{k=1}^\infty  \frac{1-|\rho_k|}{k} \left[1-z_k\right]\nn\\
&\equiv& \log Z_1+\log Z_2.\label{eq.Smin}
\end{eqnarray}
Here in the first line some simplification has been done with the properties of the exact solution. The first part is known since we have the explicit expressions for all the single-particle partition functions, whose high-temperature expansions read
\begin{eqnarray}
z_{v}(q)&=& \frac{4R^3}{\beta^3} - \frac{2R}{ \beta} + 1 - \frac{11 \beta}{60R }+{\mathcal O}(\beta^3),\label{eq.zv-exp}\\
z_{s}(q)&=& \frac{2R^3}{\beta^3} - \frac{\beta}{120R }+{\mathcal O}(\beta^3),\label{eq.zs-exp}\\
z_{f}(q)&=& \frac{4R^3}{\beta^3} - \frac{R}{2\beta} + \frac{17\beta}{480R }+{\mathcal O}(\beta^3)\label{eq.zf-exp}.
\end{eqnarray}
In particular, the leading behavior is
\begin{equation}
z(q)\to 2~ {\mathcal N}^{dof} ~\beta^{-3},
\end{equation}
with ${\mathcal N}^{dof}$ the number of independent degrees of freedom of each field. Summing them up, we get the first part in (\ref{eq.Smin})~\cite{Aharony:2003sx,Zuo:2014yga}
\begin{equation}
\frac{\log Z_1}{N^2}=(4+2 n_s+\frac{7}{2}n_f)~\zeta(4)~\frac{R^3}{\beta^3}-(2+\frac{1}{4}n_f)~\zeta(2)~\frac{R}{\beta}+{\mathcal O}(\beta^3).\label{eq.S-HT}
\end{equation}
Such an expansion corresponds to completely deconfined colored fields, which has previously been derived using the heat-kernel method~\cite{Burgess:1999vb}. Notice that the linear term in $\beta$, when combined with the previously omitted Casimir energy in the path integral, vanishes exactly~\cite{Zuo:2014yga}. Vanishing of the ${\mathcal O}(\beta)$ term also occurs in the confining phase, which is discussed in the next section.

At finite temperature the colored fields are combined by temperature-dependent $\rho_k$'s to satisfy the Gauss's law.  Away from infinite temperature, it is found that the Fourier modes $\rho_k$'s deviate from unity as $\beta^3$. The combination of all these corrections gives a finite constant to the effective action, instead of a logarithmical divergence~\cite{Zuo:2014yga}. Explicitly, the solution of the density distribution $\rho(\theta)$ will have nontrivial support in $[-\theta_0, \theta_0]$ with $0\le\theta_0<\pi$. Defining $s^2\equiv \sin ^2 (\theta_0/2)$, the asymptotic solutions are specified by
\begin{eqnarray}
s^2&=&1/\sum_{k=1}^\infty 2k z_k+{\mathcal O}(s^4)\nonumber\\
\rho_k&=& 1-\frac{k^2}{2} s^2+{\mathcal O} (s^4).\label{eq.rho-s}
\end{eqnarray}
Substituting them into (\ref{eq.Smin}), one finds a very simple correction
\begin{equation}
\log Z_2 = -\frac{N^2}{4}+{\mathcal O}(s^2).\label{eq.S-corr}
\end{equation}
The omitted terms are now at least of ${\mathcal O}(\beta^3)$, since $s^2\sim \beta^3$. (\ref{eq.S-HT}) and (\ref{eq.S-corr}) then give the complete high-temperature expansion up to ${\mathcal O}(\beta^3)$. Note that the derivation does not depend on the explicit form of $z_k$, or the matter content. Therefore the above results will be generally valid as long as $z_1\to \infty$ when $T\to \infty$. This is equivalent to the condition such that the Hagedorn transition occurs. For example, if non-trivial $R$-symmetry chemical potentials are included~\cite{Yamada:2006rx}, the result (\ref{eq.S-corr}) remains unchanged.  It is also valid with the presence of fundamental matter~\cite{Schnitzer:2004qt}, since the latter only introduces single-traces of $U^k$ in (\ref{eq.Z}), which are suppressed in the large-$N$ limit.

Here we want to make a comparison of the correction term (\ref{eq.S-corr}) with those observed in lattice simulation in flat space-time~\cite{Boyd:1996bx,Panero:2009tv}. Similar to the sub-leading term in (\ref{eq.S-HT}), the lattice simulation gives an entropy which deviates from the leading extensive term by a quadratic correction in the high-temperature expansion. An intuitive explanation attributes such a term to the residual effects of a confining string tension~\cite{Pisarski:2006yk,Narison:2009ag,Zuo:2014oma}, in addition to the rigid bag effect~\cite{Chodos:1974je}. The difference from the correction (\ref{eq.S-corr}) could be due to the different coupling regime. In the weakly-coupling regime, the effect of the confining string tension is completely invisible, and only the global gauge-invariant constraint remains, leading a constant reduction of the entropy~\cite{Zuo:2014yga}.

\subsection{Entanglement entropy and surface contribution}
Now we try to show how such a thermal entropy is related to the entanglement entropy in a $4+1~D$ gauge theory. The reasoning will be completely parallel to the corresponding $U(1)$ case recently studied in~\cite{Pretko:2015}. So we will be quite brief, and mainly focus on the differences from that in~\cite{Pretko:2015}. To be specific, we consider the deconfined phase of a pure $U(N)$ gauge theory in a flat spacetime. Since no matter fields are explicitly introduced, the theory could only have closed loops of gauge flux, in order to be gauge invariant. We want to calculate the entanglement entropy of such a phase in the simplest case, when the entangling surface is the 3-sphere. Due to the nonlocal nature of the vacuum state, such a calculation could be quite involved. It turn out to be helpful if we define the theory on a lattice, and take the continuum limit in the end. This could always be done if we we are working in the weak-coupling regime. The total wave-function can be specified by all the link variables. After the partition, the reduced density matrix will depend on the links both inside and on the entangling surface. This results in two different contributions to the entanglement entropy, the bulk and the surface contributions. The surface contribution further consists of different sectors, with each sector specified by the irreducible representations of edge links. Although the formalism has been intensively studied and formally established, a concrete calculation for a specific theory is still complicated. Only for discrete gauge theories, there are finite boundary sectors and the summation could be explicitly carried out. In the terminology of string-net condensation~\cite{Levin:2004mi}, we call such a formalism the ``string-net'' picture.

String-net condensation also has another picture, the ``close-string'' picture~\cite{Wen2004}, which seems to be more physical for gauge theories. In~\cite{Pretko:2015} such a picture is applied to the compact $U(1)$ theory in the deconfined phase. The phase is then a closed-string liquid. After the partition, the closed strings are cut into open ones, with the endpoints representing elementary charge excitations. In this way the entanglement entropy is decomposed into the charge contribution on the surface and the gauge contribution in the bulk. According to the Bisognano-Wichmann theorem~\cite{Bisognano:1976za}, the surface contribution is represented by a thermal gas of charges at infinite temperature, while the bulk contribution is summarized through a bulk integral with position-dependent temperatures. The exact relation between the surface temperature and the mass gap in the fundamental theory is not really needed. The scaling behavior of the entanglement entropy and the topological term will not depend on such details. In particular, the topological term is determined by the charge-neutral condition on the entangling surface, which in the $U(1)$ case is argued to take the form of the logarithm of boundary link number~\cite{Pretko:2015}.

This procedure could be immediately generalized to the non-Abelian case. The elementary ``charge'' excitations in the pure $U(N)$ gauge theory are of course the gluons, forming the adjoint representation. Our results (\ref{eq.S-HT}) and (\ref{eq.S-corr}), with $n_s=n_f=0$, then represent the surface contribution to the entanglement entropy of the non-Abelian theory. If the pure gauge theory is emergent from a local system with also ``colored'' open strings, one could even relate the surface temperature to the mass gap of these excitations. Independent of such details, one notice that the surface contribution naturally satisfies the area law, together with a universal topological term, $-N^2/4$. One may think that the area law should be trivially satisfied. In the next section we will show that this is not always true.

To confirm that the constant term in (\ref{eq.S-corr}) is indeed topological, we need to show that it is independent of the shape of the entangling surface. Such a proof is immediate, since details of the surface only change the single-particle partition functions. As we have emphasized, different forms of the single-particle partition functions do not change (\ref{eq.S-corr}).
The only condition we have used in the derivation of (\ref{eq.S-corr}) is that $z_1\to \infty$ at infinite temperature. Or equivalently, the Hagedorn transition must occur. This is of course necessary in order that we are dealing with the deconfined phase. Later we will give an explicit example, namely the deformed theory on the orbifold $\mathbb{S}^3/\mathbb{Z}_n$, to show that the same term remains. Finally we would like to mention that even turning on nonzero $n_s$ and $n_f$ does not change the topological term. This could also be easily understood. These adjoint matters are simply added to the already existing closed loops, therefore do not modify the close-loop constraint. Similar situation was found in the case of the $Z_2$ gauge theory coupled with fermions~\cite{Yao:2010}.

\subsection{Dimensional reduction and geometric entropy}
Next we turn to the second point of view, and study EE in the dimensional reduced theory defined through the high-temperature limit. Let us take the pure gauge theory as an example. In the high-temperature limit, the $4D$ gauge field is simply decomposed into a $3d$ vector, together with a scalar from the holonomy along the time circle. Thus we are left with a Euclidean theory of these two fields on $\mathbb{S}^3$. In the deconfined phase, these fields must again satisfy the Gauss's law. One may attempt to write the explicit lagrangian of the reduced theory. But how to implement Gauss's law constraint in $3D$? The only way known to us is through the high-temperature limit of the matrix integral (\ref{eq.Z}). Notice that in such a formalism the scalar is always needed, whose zero mode on $\mathbb{S}^3$ becomes the matrix $U$. But more matter fields, such as the fermions and the conformal scalars, do not introduce any further complexity.

In such a Euclidean theory, EE could be naturally defined through the replica trick, and named geometry entropy~(GE)~\cite{Callan:1994py}. In particular, as emphasized in~\cite{Fujita:2008zv}, GE naturally generalizes the notion of TEE in a topological field theory, such as the Chern-Simons gauge theories on $\mathbb{S}^3$~\cite{Dong:2008ft}. Since our 3D theory can only be defined with the extra dimension, we switch the order of the high-temperature limit and the replica trick. That means, one first generalizes the concept of the geometry entropy to finite temperature~\cite{Fujita:2008zv}. Taking the high-temperature limit in the end together with proper subtraction of the divergent terms, one should be able to recover GE in the reduced theory.

The derivation of GE in the deconfined phase has been partially done in~\cite{Fujita:2008zv}. Here we want to improve it to include the sub-leading term, which will be the only nontrivial term after the dimensional reduction. To introduce GE, one first expresses the metric of $\mathbb{S}^3$ as
\begin{equation}
\mathd \Omega_3^2=\mathd \theta^2+ \sin ^2\theta (\mathd \psi^2+\sin^2\psi \mathd \phi^2),
\end{equation}
with $0\le\theta,\psi\le \pi$, and $0\le \phi\le 2\pi$. In order to apply the replica method, one defines the partition function $Z(l)$ on the $l$ copy of the original spacetime, with $0\le \phi\le 2\pi l$. Then GE can be computed through~\cite{Fujita:2008zv}
\begin{equation}
S_G =-\frac{\partial}{\partial l}~\log \left[ \frac{Z(l)}{Z(1)^l} \right]\Big|_{l=1} .\label{eq.SG}
\end{equation}
For $\mathbb{S}^3$, it is more convenient to take $l=1/n$ with $n$ an integer. In this case the auxiliary spacetime becomes the orbifold $\mathbb{S}^3/\mathbb{Z}_n$. The previous partition function on $\mathbb{S}^3$ (\ref{eq.Z}) can be directly generalized to the orbifold case, by replacing the single-particle partition functions with the corresponding ones on the orbifold.
These are very neatly derived in the appendix of~\cite{Fujita:2008zv}:
\begin{equation}
z^{(n)}_v(q)=\frac{2q^2(1+2q^{n-1}-q^n)}{(1-q)^2(1-q^n)},\quad z^{(n)}_s(q)=\frac{q(1+q^n)}{(1-q)^2(1-q^n)}, \quad z^{(n)}_f(q)=\frac{4 q^{1+n/2}}{(1-q)^2(1-q^n)}.\label{eq.z-n}
\end{equation}
When $n\to 1$ they recover (\ref{eq.z-single}).
Their high-temperature expansion are as follows
\begin{eqnarray}
z^{(n)}_{v}(q)&=& \frac{4R^3}{n\beta^3} + \frac{(n^2-6n-1)R}{ 3n\beta} + 1 +{\mathcal O}(\beta),\label{eq.znv-exp}\\
z^{(n)}_{s}(q)&=& \frac{2R^3}{n\beta^3} +\frac{(n^2-1)R}{ 6n\beta}+{\mathcal O}(\beta),\label{eq.zns-exp}\\
z^{(n)}_{f}(q)&=& \frac{4R^3}{n\beta^3} - \frac{(n^2+2)R}{ 6n\beta} +{\mathcal O}(\beta)\label{eq.znf-exp}.
\end{eqnarray}
Notice that the leading term is proportional to the volume of $\mathbb{S}^3/\mathbb{Z}_n$ , and thus always $1/n$ of those in (\ref{eq.zv-exp}, \ref{eq.zs-exp}, \ref{eq.zf-exp}). As a result, they do not contribute to GE as we will see. The effective action can be derived in the same way as in (\ref{eq.Smin}). The deconfined part from the individual single-particle states are
\begin{eqnarray}
\frac{\log Z_1^{(n)}}{N^2}&=&(4+2 n_s+\frac{7}{2}n_f)~\zeta(4)~\frac{R^3}{n\beta^3}\nonumber\\
\quad                   &&+\left[ \frac{(n^2-6n-1)}{ 3n} +\frac{(n^2-1)}{ 6n}n_s - \frac{(n^2+2)}{ 12n} n_f \right]\frac{\zeta(2)R}{\beta}+{\mathcal O}(\beta).\label{eq.Sn-HT}
\end{eqnarray}
Notice that for $n\ne1$ the linear term in $\beta$ is no longer vanishing. The correction term due to the Gauss's constraint, or deviations of $\rho_k^{(n)}$ from unity, can be obtained from the corresponding saddle-point solution. In fact, we do not really need to repeat the derivation again, since in our previous results (\ref{eq.rho-s}) we do not really specify the detailed form of $z_k$. Therefore, we find the same correction independent of $n$
\begin{equation}
\log Z_2^{(n)} = -\frac{N^2}{4}+{\mathcal O}((s^{(n)})^2),\label{eq.Sn-corr}
\end{equation}
where the corrections are also of ${\mathcal O}(\beta^3)$. Now we can combine all the results, to obtain GE from (\ref{eq.SG})
\begin{equation}
S_G=\left[-\frac{4}{3}+\frac{n_s}{3}-\frac{n_f}{6}\right]
\frac{\zeta(2)N^2R}{\beta}-\frac{N^2}{4}+{\mathcal O}(\beta).\label{eq.SG-exp}
\end{equation}
Indeed the volume term, proportional to $R^3$, vanishes. One can check that the coefficients of the leading term in the above expression agree with~\cite{Fujita:2008zv}. After subtracting the divergent ${\mathcal O}(\beta^{-1})$ term, only the constant term, $\gamma=-N^2/4$, remains in the reduced 3D theory. According to the argument given in~\cite{Fujita:2008zv}, we conclude that the reduced theory contains some nontrivial topological structure with a nontrivial TEE $\gamma$. This could be thought of as a generalization of the calculation of TEE in Chern-Simons theories~\cite{Dong:2008ft}.

Let us discuss a little more for the special case, ${\mathcal N}=4$ super Yang-Mills theory~(SYM). The matter content is specified by $n_s=6$ and $n_f=4$. One can check that the leading term in (\ref{eq.SG-exp}) vanishes~\cite{Fujita:2008zv}. So GE is immediately given by a single constant at $\beta=0$
\begin{equation}
S_G^{\rm{SYM},\lambda\to 0}=-\frac{N^2}{4}.
\end{equation}
 At strong coupling the geometric entropy can be calculated through the gauge/string duality. In the holographic approach the thermodynamics for ${\mathcal N}=4$ SYM on $\mathbb{S}^3$ has been studied intensively with the corresponding black hole background~\cite{Hawking:1982dh,Witten:1998zw}\cite{Burgess:1999vb,Zuo:2014vga}. A direct generalization to $\mathbb{S}^3/\mathbb{Z}_n$ gives~\cite{Fujita:2008zv}
\begin{eqnarray}
\frac{S_G^{\rm{SYM},\lambda\to \infty}}{N^2}&=&-\frac{\pi^2 R}{8\beta} \left( 1+\sqrt{1-\frac{2\beta^2}{\pi^2 R^2}}  \right)^2\nonumber\\
&=&-\frac{\pi^2 R}{2\beta}+\frac{\beta}{2R}+{\mathcal O}(\beta^3),
\end{eqnarray}
where the contribution of the thermal gas phase has been subtracted. The above result gives a vanishing TEE for the reduced theory. Similar holographic calculation on a different background also gives a vanishing result~\cite{Pakman:2008ui}. This indicates that the black hole phase at strong coupling and the deconfined phase at zero coupling correspond to different phases of ${\mathcal N}=4$ SYM. Since the latter contains nontrivial topological structure after dimensional reduction, the two phases may not be interpolated continuously as coupling varies~\cite{Li:1998kd,Gao:1998ww}\cite{Zuo:2014yga}.

All the above derivation is done on $\mathbb{S}^3$. If the geometric entropy $\gamma$ is indeed topological, it should again be independent of the shape of the manifold. Same as in the previous subsection, this is again due to the fact that the derivation of (\ref{eq.Sn-corr}) does not depend on the detailed functions (\ref{eq.z-n}). As long as the theory is in the deconfined phase, all these single-particle partition functions would diverge in the high temperature limit. This makes  possible the derivation of (\ref{eq.Sn-corr}) through a high-temperature expansion. However, there is one more necessary condition here for the three manifold. One needs at least one periodic direction, $0\le \phi\le 2\pi$ here, for the geometric entropy to be properly defined. When such a condition is satisfied, the geometric entropy could be interpreted as the entanglement entropy on a spatial section of fixed $\phi$.

\section{Confining phase}
\label{sec:ConP}
\subsection{Thermal entropy and 2D modular formulae}
 For comparison, we also apply the same procedure to the confined phase, to see if similar topological term could be extracted. According to the analyzes in ~\cite{Sundborg:1999ue,Aharony:2003sx}, this phase is characterized by a uniform distribution of the eigenvalues, resulting a vanishing effective action at ${\mathcal O}(N^2)$. The nontrivial effective action at ${\mathcal O}(N^0)$ can be directly derived through integration of the fluctuations around the uniform saddle point. Alternatively, it can be derived by counting the colorless states at a given energy. In this way one has to construct the single-trace partition function from the single-particle ones, taking into account the cyclic symmetry of the trace operation. Employing the P\'{o}lya enumeration theorem, the finally result is~\cite{Sundborg:1999ue,Polyakov:2001af,Aharony:2003sx}
\begin{equation}
Z_{ST}(q)=-\sum_{k=1}^\infty \frac{\phi(k)}{k}\log (1-z_k),\label{eq.Z-ST}
\end{equation}
where $\phi(k)$ is the Euler totient function, and $z_k$ is defined as (\ref{eq.zk}). The full partition function should contain the contributions of all the multi-trace states, which are not independent at finite $N$. In the $N\to \infty$ limit the constraints among different multi-trace operators are relaxed, and one obtains a simple expression
\begin{equation}
\log Z(q)=\sum_{m=1}^\infty \frac{Z_{ST}(q^m)}{m}=-\sum_{k=1}^\infty \log \left[ 1-z_k(q)\right],~~~Z=\prod_{k=1}^\infty~\frac{1}{1-z_k(q)}. \label{eq.Z-full}
\end{equation}
In other words, $Z$ is the plethystic exponential of $Z_{ST}$~\cite{Benvenuti:2006qr}. One could also perform the inverse operation, the plethystic logarithm, to recover $Z_{ST}$ from $Z$~\cite{Benvenuti:2006qr}.
\begin{equation}
Z_{ST}(q)=\sum_{n=1}^\infty \frac{\mu(n)}{n}~\log Z(q^n),
\end{equation}
with $\mu(n)$ the M\"{o}bius function.

The above partition function (\ref{eq.Z-full}) develops an instability when the argument of the logarithm vanishes. The instability is induced by the exponentially increasing of the density of states with a fixed energy, which is called Hagedorn instability~\cite{Hagedorn:1965}. A Hagedorn transition occurs at the singular point and the theory goes to the deconfined phase discussed before.

For comparison, we also include the corresponding expression for the case when the fermions satisfy ``twisted'' boundary condition~\cite{Kovtun:2007py,Unsal:2007fb}. That is, they are periodic along the Euclidean time circle. The full twisted partition function can be similarly constructed~\cite{Basar:2014jua}
\begin{equation}
\log \tilde Z=-\sum_{k=1}^\infty \log \left[ 1-\tilde z_k  \right], ~~~\tilde Z=\prod_{k=1}^\infty~\frac{1}{1-\tilde z_k(q)} \label{eq.tZ-full}
\end{equation}
where $\tilde z_k\equiv z_v(q^k)+n_sz_s(q^k)-n_fz_f(q^k)$. Due to the cancellation between the bosonic and fermionic contributions, the Hagedorn instability may disappear~\cite{Unsal:2007fb}. For the special case without scalars, a nonzero $n_f$ removes the instability~\cite{Basar:2014jua}. In Appendix \ref{App-twist} we determine the general condition of $n_s$ and $n_f$ for the instability to appear.

\subsubsection{Naive high-temperature expansion}
We try to take the high-temperature expansion of (\ref{eq.Z-full}) again. First we adopt a naive procedure to see the general expansion pattern, as done in~\cite{Basar:2014jua,Basar:2014hda}~\footnote[2]{This procedure is later employed to derive spectral sum rules in extended theories~\cite{Cherman:2015rpc}.}. For completeness, we give a short derivation of the expansion scheme.

The single-particle partition function $z_f$ for fermions involves half-integer powers of $q$. It is then convenient to define
 \begin{equation}
 Q\equiv q^{1/2}.
 \end{equation}
The combination of the single-particle functions can then be expanded as
\begin{eqnarray}
1-z_k&=&\frac{Q^{6k}-(3+n_s)Q^{4k}+4n_f(-Q)^{3k}-(3+n_s)Q^{2k}+1}{(1-Q^{2k})^3}\nonumber\\
                                        &\equiv& \frac{P((-Q)^k)}{(1-Q^{2k})^3}.\label{eq.z-combination}
\end{eqnarray}
To be more specific, in the above equation we have defined the polynomial $P(x)$ as
\begin{equation}
P(x)\equiv x^6-(3+n_s)x^4+4n_f x^3-(3+n_s)x^2+1.\label{eq.Px}
\end{equation}
To get rid of the logarithm in (\ref{eq.Z-ST}) and (\ref{eq.Z-full}), we take the logarithmic derivative with respect to $Q$. The individual term in the sum (\ref{eq.Z-ST}) then possesses the asymptotic expansion when $Q\to 1$:
\begin{eqnarray}
Q\frac{\partial}{\partial Q} \log\left[ 1-z_k\right]   &=&\frac{Q \frac{\partial}{\partial Q}P((-Q)^k)}{P((-Q)^k)}+\frac{6kQ^{2k}}{1-Q^{2k}}\nonumber\\
&  =&3k+\frac{6kQ^{2k}}{1-Q^{2k}}.
\end{eqnarray}
Here the first term has been evaluated at $Q=1$, since both the denominator and numerator are non-vanishing there~\footnote[3]{For special matter content it could happen $P(1)=0$, but this does not change the naive expansion here. See Appendix \ref{App-twist} for more details.}. Substituting this into (\ref{eq.Z-ST}) and finishing the summation with the properties of $\phi(k)$, one finds~\footnote[4]{Here a naive continuation of the Dirichlet series of the Euler totient function has been used~\cite{Basar:2014hda}. A rigorous treatment is later supplemented in~\cite{Cherman:2015rpc}.}
\begin{equation}
Z_{ST}'(\beta)=\frac{3R}{\beta^2}+{\mathcal O}(\beta^2),
\end{equation}
where $'$ denotes the derivative with respect to $\beta$.
Obviously in such a regularization the constant term in $Z_{ST}$ is lost. This can be recovered by expanding directly the logarithm
\begin{eqnarray}
\log\left[ 1-z_k \right]&=&-3\log k + \log\left[4+2n_s-3(-1)^kn_f\right]\nonumber\\
&&-3\log[q-1]+\frac{3}{2}(q-1)+{\mathcal O} ((q-1)^2).
\end{eqnarray}
Only the first term contributes nontrivially to the summation (\ref{eq.Z-ST}), resulting a constant correction to $Z_{ST}$
\begin{equation}
\delta Z_{ST} (\beta) = \frac{3}{2}.
\end{equation}
Combining the two pieces, one gets
\begin{equation}
 Z_{ST}(\beta)=-\frac{3R}{\beta}+\frac{3}{2}+{\mathcal O}(\beta^3).\label{eq.ZST-ex}
\end{equation}
Further substituting this in (\ref{eq.Z-full}), the full partition function acquires a similar expansion
\begin{equation}
\log Z(\beta)=-\frac{3R}{\beta}\zeta(2)+\frac{3}{2}\zeta(1)+{\mathcal O} (\beta^3).\label{eq.Z-reg}
\end{equation}
The same procedure can be applied to the twisted expression (\ref{eq.tZ-full}), though some care must by taken on the poles/zeroes. Astonishingly, the volume terms of ${\mathcal O}(\beta^{-3})$ in the expansion (\ref{eq.zv-exp},\ref{eq.zs-exp},\ref{eq.zf-exp}) of the single-particle partition functions completely disappear. As emphasized in \cite{Basar:2015xda} and recently in~\cite{Basar:2015asd}, this reflects some kind of $4D-2D$ equivalence. Here we want to argue that, disappearance of the volume term is due to the strong colorless constraints in constructing the single-trace partition function (\ref{eq.Z-ST}). These constraints are so strong that no free-propagating modes in the total space appear any more. Due to this, no reference entropy, with the volume scaling, exists for the entropy deficit to be unambiguously defined. Another difference from (\ref{eq.S-corr}) is the ${\mathcal O}(\beta^0)$ term, which is now divergent as $\zeta(1)$, instead of being finite. Such a divergence comes from the infinite sum of the multi-trace states, which is guaranteed only in the large-$N$ limit. At finite $N$ extra constraints appear and the number of multi-trace states are bounded from above. We also notice that the above expansion is universal and completely independent of the matter content. An immediate question is, in what sense would this be true?

\subsubsection{2D modular formulae and high-temperature continuation}\label{subsec:2D}
Although the above procedure illustrates the expansion pattern and shows clearly the physical picture, it can not give exact results simply because of the existence of the Hagedorn singularities on the real line $T>0$. For such a case a proper formulation will be, going to the complex plane. We thus define
\begin{equation}
\tau\equiv \mathi \beta/(4\pi R), \quad \mbox{i.e.} \quad  \mathe ^{2\mathi \pi \tau}\equiv Q=q^{1/2},
\end{equation}
and treat $\tau$ as a complex variable. The high-temperature expansion then corresponds to the small-$|\tau|$ expansion. Still, the infinite product in (\ref{eq.Z-full}) and (\ref{eq.tZ-full}) makes the expansion not easy to apply. Some recent progresses reveal that this infinite product can be neatly performed, giving a compact expression for the full partition function in terms of toric modular functions~\cite{Basar:2014jua,Basar:2015xda}. The final expressions for some special cases are presented there, but the generalization to arbitrary matter content, with thermal or twisted boundary conditions, are needed for our purpose. So we perform a short derivation for the general expressions, and then use them to derive the small-$|\tau|$ expansion~\footnote[5]{The modular expressions in general situation is discussed in detail later~\cite{Basar:2015asd}.}.

As in~\cite{Basar:2014jua}, we start with the twisted partition function (\ref{eq.tZ-full}). Similar as (\ref{eq.z-combination}), in the twisted case one finds
\begin{eqnarray}
1-\tilde z_k&=&\frac{Q^{6k}-(3+n_s)Q^{4k}+4n_f~Q^{3k}-(3+n_s)Q^{2k}+1}{(1-Q^{2k})^3}\nonumber\\
                                        &=& \frac{P(Q^k)}{(1-Q^{2k})^3}.\label{eq.tz-combination}
\end{eqnarray}
This shows clearly that under $Q\to  -Q$,
\begin{equation}
\tilde z_k \to  z_k,~~~  \tilde Z\to  Z.
\end{equation}
So any expression of $\tilde Z$ can be immediately converted to $Z$ through $Q\to  -Q$, or $\tau \to \tau +1/2$. In order to perform the infinite product, we need to analyze in detail the roots of $P(Q)=0$.

An important ingredient that simplifies the root structure, is the so-called ``temperature-reflection symmetry" found in~\cite{Basar:2014mha}. Under such a symmetry transformation, the partition function in some systems is symmetric up to a multiplicative factor. An immediate inference will be vanishing of vacuum energy, since it is extracted from the linear term ${\mathcal O}(\beta)$ in the high-temperature expansion of $\log Z(\beta)$. In the present case such a symmetry is also valid~\cite{Basar:2014mha}. This is most manifest in (\ref{eq.z-combination}) and (\ref{eq.tz-combination}), which is symmetric under $\beta \to -\beta$, or $Q\to 1/Q$. Due to this, the roots of the equation, $P(Q)=0$, come in reciprocal pairs. Denoting the roots as $Q_i=-r_i, Q_{i+3}=-1/r_i$ for $i=1,2,3$, the polynomial $P(Q)$ can be factorized as
\begin{equation}
P(Q)=\prod_{i=1}^3 (1+r_i Q)(1+r_i^{-1} Q).\label{eq.PQ}
\end{equation}
This can be easily generalized to $ P(Q^k) $. As a result, the full twisted partition function has the product expression
\begin{eqnarray}
\tilde Z&=&\prod_{k=1}^\infty\frac{1}{  1-\tilde z_k }\nonumber\\
&=&\prod _{k=1}^\infty \prod _{i=1}^3 \frac{1-Q^{2k}}{ (1+r_i Q^k)(1+r_i^{-1} Q^k)}.
\end{eqnarray}
Very interestingly, the infinite product can be compactly expressed through the Dedekind $\eta$-function and Jacobi $\vartheta$-functions. Generalizing the derivation in the appendix of~\cite{Basar:2014jua}, one obtains a delicate formula
\begin{equation}
\tilde Z(\tau)=\eta^3(2\tau)  \prod _{i=1}^3 \left[ \frac{2\cos (\pi \nu_i)~\eta(\tau)}{\vartheta_2(\nu_i| \tau)}    \right],\label{eq.tZ-tau}
\end{equation}
where $\mathe ^{2\mathi \pi \nu_i}\equiv r_i$.
Note that the Dedekind $\eta$-function is defined as
\begin{equation}
\eta(\tau)=\mathe ^{\frac{\mathi \pi \tau}{12}} \prod _{n=1}^\infty (1-\mathe ^{2\mathi \pi \tau n}),
\end{equation}
and the notation for the second Jacobi $\vartheta$-function is
\begin{equation}
\vartheta_2(\nu|\tau)=2\xi ^{1/4} \cos (\pi \nu) \prod _{n=1}^\infty (1-\xi^{2n})(1+2~\cos (2\pi \nu)~\xi ^{2n}+\xi^{4n}),
\end{equation}
with $\xi\equiv \mathe ^{\mathi \pi \tau}$. Similar formulae are frequently employed in supersymmetric theories, see e.g., ~\cite{Bourdier:2015sga}. Finally through the translation $\tau\to \tau +1/2$, the thermal partition function takes the form
\begin{equation}
Z(\tau)=\eta^3(2\tau+1)  \prod _{i=1}^3 \left[ \frac{2\cos (\pi \nu_i)~\eta(\tau+\frac{1}{2})}{\vartheta_2(\nu_i| \tau+\frac{1}{2})}    \right],\label{eq.Z-tau}
\end{equation}
It can also been expressed through generalized $\vartheta$-functions, for pure Yang-Mills theory~\cite{Basar:2015xda} and in general~\cite{Basar:2015asd}. It is further shown that the modular structure originates from some $2D$ irrational CFT~\cite{Basar:2015xda,Basar:2015asd}. As already noticed in~\cite{Basar:2014jua}, both (\ref{eq.Z-tau}) and (\ref{eq.tZ-tau}) are invalid when $\nu_i=1/2$, or $r_i=-1$. In the main text of this paper we always assume $r_i\ne-1$. The situation $r_i=-1$ will be discussed in Appendix~\ref{App-twist}. It will be shown that the partition function can still be expressed through the modular functions. An interesting question is whether the single-trace partition function could also be simplified by applying the plethystic log to the above modular expression~\cite{Benvenuti:2006qr}.

Once the partition function is expressed through the modular functions, we can extract its properties with the powerful techniques of modular transformations. Here we simply need the small-$|\tau|$ expansion, which can be easily derived through the $S:~\tau\to -1/\tau$ and $T:~\tau\to \tau +1$ transformations. There is only one additional complexity here. The argument of the $\vartheta$ functions is $\tau +1/2$ instead of $\tau$. The behavior of $\vartheta$ near $\tau = 1/2$ can be derived with the help of the Landen transformation~\cite{Wang1989}. Putting all the pieces together, one finds the following expansion:
\begin{equation}
\log Z(\tau)=-A\frac{\mathi \pi}{\tau}-\frac{3}{2}\log(-\mathi \tau) +C+{\mathcal O}(\mathe^{-\mathi /\tau}),\label{eq.Ztau-exp}
\end{equation}
where $A$ is a function of $\nu_1,~\nu_2,~\nu_3$, and the residual part, denoted as $C$, contains the oscillation terms and constant terms. The above expansion can be unambiguously fixed by first taking the small-$|\tau|$ expansion of $Z(\tau)$, and then $\arg \tau\to \pi/2$.
In such a way, $A$ is guaranteed to be real. The explicit form of $A$ and $C$ depends on the positions of the roots, and we do not attempt to derive them in general~\footnote[6]{A generic expression for $A$ is later derived by carefully analyzing the infinite sums involving the Euler totient function~\cite{Cherman:2015rpc}.}. In the simple case of pure Yang-Mills theory, the expansion can be expressed as
\begin{equation}
\log Z_{YM}(\tau)=-\left(\frac{1}{16}+\frac{2\theta^2}{\pi^2}\right)\frac{\mathi \pi}{\tau}-\frac{3}{2}\log (-\mathi \tau)-\log \left( \sin \frac{\mathi \theta}{\tau}\right)+\log (\cosh \theta \sinh \theta)+{\mathcal O}(\mathe^{-\mathi /\tau}),\label{eq.ZYM-exp}
\end{equation}
where $\theta=-\log (2+\sqrt 3)/4$. In Appendix~\ref{App:YM} we derive this from both (\ref{eq.Z-tau}) and (\ref{eq.tZ-tau}), which give essentially the same result.
After the expansion (\ref{eq.Ztau-exp}) is obtained, we can substitute $\tau=\mathi \beta/4\pi R$ and get
\begin{equation}
\log Z(\beta)=-A\frac{4\pi^2R }{\beta}-\frac{3}{2}\log\frac{ \beta}{4\pi R} +C+{\mathcal O}(\mathe^{-4\pi R /\beta}).\label{eq.Z-beta}
\end{equation}
We notice that similar expressions also appear in weak-coupling calculation on a different manifold~\cite{Zhitnitsky:2013hs,Zhitnitsky:2013hba}. For pure Yang-Mills theory $A=\frac{1}{16}+\frac{2\theta^2}{\pi^2}>0$, and the total coefficient of the leading term is negative. A reasonable guess would be that, the leading term will always be negative when the Hagedorn instability is present.

The above expansion confirms and generalizes the main structure of the naive expansion (\ref{eq.Z-reg}). Actually, it tells us much more. First, it confirms that the volume term is indeed vanishing. Secondly, the divergence at ${\mathcal O}(\beta^0)$ turns out to be a logarithmic divergence in $\beta$. The coefficient of this term is indeed universal, in the sense that it does not depend on particular values of $n_s$ and $n_f$. Only in the special case when $P(Q=1)=0$ for some matter content, the coefficient changes according to the rate $P(1)$ vanishes~(see Appendix~\ref{App-twist}). However, the coefficient of the leading term, $A$, and the form of the residual term do depend on the matter content. Furthermore, the above expansion stops at ${\mathcal O}(\beta^0)$, up to exponentially suppressed terms. This follows the famous pattern of the corresponding expansion in 2D CFT~\cite{Cardy:1986ie}. In particular, the linear term in $\beta$, and thus the vacuum energy, vanishes~\cite{Basar:2014hda,Basar:2015xda}. Recent development in high-temperature expansion of 4D supersymmetric partition functions shows very similar structure as the above expression, see~\cite{DiPietro:2014bca,Ardehali:2015hya}.

\subsection{Entanglement entropy and surface contribution}
Now we follow the same reasoning as in the deconfined phase, to interpret the high temperature limit of the thermal entropy as the surface contribution of the entanglement entropy. It is anticipated in \cite{Pretko:2015} that in the confined phase of a $U(1)$ theory, the area-law term of the entanglement entropy will not be present. Here we have shown that, at least for the surface contribution, the area-law term is always missing in both the naive estimation (\ref{eq.Z-reg}) and the exact expansion (\ref{eq.Z-beta}). As argued in~\cite{Pretko:2015}, this is due to the fact that the allowable boundary conditions are not fully sampled by the corresponding wave function. In other words, the boundary particles are not allowed to propagate freely. Accordingly, the Gauss's law is built into the local degrees of freedom, and no topological term will be present. Practically, the appearance of the logarithmic divergence in (\ref{eq.Z-beta}) makes the extraction of a universal constant term impossible. It was argued to be due to the possible fractional modes of order $1/(N\beta)$ when the Polyakov loop is trivial~\cite{Aharony:2005ew}. Some explanations for the appearance of the logarithmic divergence in supersymmetric theories are given in~\cite{DiPietro:2014bca,Ardehali:2015hya}.


\subsection{Dimensional reduction and geometric entropy}

Finally we want to check the above conclusion from the view point of dimensional reduction, by calculating the geometric entropy. With the single-particle partition functions (\ref{eq.z-n}), we could easily generalize the confining-phase partition functions to the orbifold $\mathbb{S}^3/\mathbb{Z}_n$
\begin{eqnarray}
Z_{ST}^{(n)}(q)&=&-\sum_{k=1}^\infty \frac{\phi(k)}{k}\log (1-z^{(n)}_k),\label{eq.Zn-ST}\\
\log Z^{(n)}&=&\sum_{m=1}^\infty \frac{Z^{(n)}_{ST}(q^m)}{m}=-\sum_{k=1}^\infty \log \left[ 1-z^{(n)}_k(q)  \right], ~~~Z^{(n)}=\prod_{k=1}^\infty~\frac{1}{1-z^{(n)}_k(q)}, \label{eq.Zn-full}
\end{eqnarray}
where
$z^{(n)}_k(q)\equiv z^{(n)}_v(q^k)+n_sz^{(n)}_s(q^k)-(-1)^kn_fz^{(n)}_f(q^k)$. From (\ref{eq.z-n}), one gets
\begin{eqnarray}
&&1-z^{(n)}_v(q^k)-n_sz^{(n)}_s(q^k)+(-1)^kn_fz^{(n)}_f(q^k)\nonumber\\
&&\quad \quad =\frac{Q^{(2n+4)k}-(2+n_s)Q^{(2n+2)k}-Q^{2nk}+4n_f(-Q)^{(n+2)k}-Q^{4k}-(2+n_s)Q^{2k}+1}{(1-Q^{2k})^2(1-Q^{2nk})}\nonumber\\
&&\quad \quad \equiv \frac{P^{(n)}((-Q)^k)}{(1-Q^{2k})^2(1-Q^{2nk})},\label{eq.zn-combination}
\end{eqnarray}
which is again invariant under $Q\to 1/Q$. So the $2n+4$ roots of $P^{(n)}(Q)=0$ also appear in reciprocal pairs. We therefore parameterize the roots as
\begin{equation}
Q_i=-r_i,~~ Q_{i+n+2}=-1/r_i ,~~\mathe ^{2\mathi \pi \nu_i}\equiv r_i, \quad i=1,2,...,n+2.
\end{equation}
Following the previous procedure, the generalized partition function $Z^{(n)}$ can be compactly expressed as
\begin{equation}
Z^{(n)}(\tau)=\eta^2(2\tau+1) \eta\left(2n\tau+n)\right)  \prod _{i=1}^{n+2} \left[ \frac{2\cos (\pi \nu_i)~~\eta(\tau+\frac{1}{2})}{\vartheta_2(\nu_i| \tau+\frac{1}{2})}    \right],\label{eq.Zn-tau}
\end{equation}
 This leads to the exact expansion
\begin{equation}
\log Z^{(n)}(\beta)=-A^{(n)}\frac{4\pi^2R }{\beta}-\frac{3}{2}\log\frac{ \beta}{4\pi R} +C^{(n)}+{\mathcal O}(\mathe^{-4\pi R /\beta}),\label{eq.logZn}
\end{equation}
in parallel with (\ref{eq.Z-beta}). Notice that the logarithmic term is independent of the replica parameter $n$.
The geometric entropy is then derived through (\ref{eq.SG})
\begin{equation}
S_G(\beta)=-\frac{\partial}{\partial (1/n)}\left[ \log Z^{(n)}(\beta)-\frac{1}{n} \log Z(\beta)  \right]\Big|_{n=1}.\label{eq.SG2}
\end{equation}
Since the logarithmic term is independent of $n$, it remains finally in $S_G(\beta)$. As a result, the geometric entropy of the dimensional reduced theory is not well defined. In other words, the logarithmic divergence makes the dimension reduction ambiguous.

Again the logarithmic divergence could be attributed to the large-$N$ limit. Applying the plenthystic log to (\ref{eq.logZn}), one finds $ Z^{(n)}_{ST}(\beta)$ contains a constant term $3/2$, in addition to the power terms. From this one could define the single-trace geometry entropy
\begin{equation}
S^{ST}_G(\beta)\equiv-\frac{\partial}{\partial (1/n)}\left[  Z^{(n)}_{ST}(\beta)-\frac{1}{n} Z_{ST}(\beta)\right]\Big|_{n=1}.
\end{equation}
After subtraction of power divergence, it becomes a universal constant term
\begin{equation}
S^{ST}_G(\beta)= \frac{3}{2}.
\end{equation}
In the large-$N$ limit, we have to sum over infinitely many of this, resulting the divergence in the full geometric entropy (\ref{eq.SG2}).

\section{Discussion}


In this paper we try to interpret the thermal entropy deficit, for deconfined $U(N)$ gauge theory on a $3D$ compact manifold, as the topological entanglement entropy in two related theories. First the relation is manifest if we identify the high-temperature limit of the thermal entropy as the surface contribution to the entanglement entropy in a $4+1 D$ bulk theory. Since the particle-like excitations on the entangling surface are deconfined, this naturally gives rise to the area-law term. The Gauss's constraint further results in the topological term. Such a derivation of the topological entanglement entropy is recently investigated in the compact $U(1)$ theory~\cite{Pretko:2015}~(see also~\cite{Radicevic:2015}). Secondly, it also coincides with geometric entropy in the dimensional reduced theory defined through the high temperature limit, generalizing the direct calculation in Chern-Simons gauge theories~\cite{Dong:2008ft}. With these relations, the thermal entropy deficit studied here could be considered as a higher-dimensional analog of the edge entropy in 2D rational CFT~\cite{Fendley:2007}. There the authors show that edge entropy is identical to the topological entanglement entropy of the bulk Chern-Simons theory, based on the same modular structure of the bulk and edge theories.

On the contrary, in the confined phase the area-law term, at least for the surface contribution considered here, is missing. As we argued previously, this reflects the fact that there is no local propagating degrees of freedom after imposing the Gauss's constraint locally. The individual physical excitations are the single-trace states, which display no local properties at all. In the $U(1)$ case, the confining phase corresponds to neutral charge everywhere on the entangling surface, also argued to lead to a vanishing area term~\cite{Pretko:2015}. Regarding such a scaling behavior, our results seems to be in accordance with the picture at strong coupling~\cite{Nishioka:2006gr,Klebanov:2007ws}.

Despite the similarity, there is a significant difference between our results and that in~\cite{Pretko:2015}. The TEE we found is finite, while that in~\cite{Pretko:2015} is logarithmic divergent as the entanglement surface size increases. According to recent studies in ~\cite{Soni:2015yga,VanAcoleyen:2015ccp}, our result corresponds to the distillable entanglement, which automatically vanishes in Abelian theories. In contrast, the contributions summed up in~\cite{Pretko:2015} correspond to the classical part, which naively diverges for a continuous gauge group. The different behavior of these two contributions indicates that the distillable entanglement would be more appropriate to extract a finite TEE. We would like to investigate this further in a revised version of \cite{Zuo:2016knh}.

 Another possible reason for the above difference could be that, we have taken the large-$N$ limit before taking the size of the entangling surface to be large. Possible sub-leading terms in the large-$N$ limit could diverge when the entangling surface increases.
When $N$ is finite, no Hagedorn transition occurs and we are left with a single ``mixed'' phase. In such a phase the effective action is roughly given by the sum of both the confining contribution with a ${\mathcal O}(N^0)$ coefficient and the deconfined one with a coefficient of ${\mathcal O}(N^2)$. Taking the high-temperature limit, one always get a logarithmic divergence from the former. Since the confined-phase spectrum coincides with that of a $2D$ irrational CFT~\cite{Basar:2015xda,Basar:2015asd}, such a logarithmic divergence is consistent with those in $2D$ CFT~\cite{Holzhey:1994we,Calabrese:2004eu}. However, it seems not easy to solve the problem at finite $N$ analytically. Some existing results in~\cite{Benvenuti:2006qr} could be of help.


The above expectation is also consistent with the results on EE in matrix quantum mechanics, see e.g., ~\cite{Calabrese:2011vh,Calabrese:2015,Hartnoll:2015fca}. In particular, it is recently suspected in~\cite{Anninos:2014ffa} that change of the collectivity of the eigenvalue distribution in matrix models will result in change of topological order. According to the fermion-gas interpretation~\cite{Brezin:1977sv}, different eigenvalue distributions correspond to different fermion surface topologies, which are intrinsic quantum orders~\cite{Wen2004}. Our analyzes show that in the deconfined phase the single-gap solution, the $A_1$-type according to the classification of~\cite{Jurkiewicz:1982iz}, already contains some nontrivial topological structure. It will be interesting to further investigate the more complicated situation classified there~\cite{Jurkiewicz:1982iz}.

\section*{Acknowledgments}
F.Z. would like to take the opportunity to express his appreciation to Ling-Yan Hung, for the stimulating discussions and nice explanations on entanglement entropy and topological order. He also thanks Zhao-Long Wang and Jun-Bao Wu for helpful discussions. Part of the work was done when F.Z. was participating the program ``Holographic duality for condensed matter physics'', held by Kavli Institute for Theoretical Physics China at the Chinese Academy of Sciences~(KITPC) in Beijing. The hospitality of KITPC is acknowledged. The work is partially supported by the National Natural Science Foundation of China under Grant No. 11405065 and No. 11445001.

\appendix

\section{Special cases with $P(Q=1)=0$}
\label{App-twist}
In the main text we have assumed that the equation $P(Q=1)\ne 0$, with $P(x)$ defined in (\ref{eq.Px}) as
\begin{equation}
P(x)\equiv x^6-(3+n_s)x^4+4n_f x^3-(3+n_s)x^2+1.\label{eq.Px2}
\end{equation}
Here we supplement the corresponding results when $P(Q=1)=0$. It is not difficult to find that, this happens when
\begin{equation}
n_f=n_s/2+1.
\end{equation}
Moreover, when $n_f\ne 4$, $Q=1$ is a double root. For convenience we set $Q_3=Q_6=-r_3=1$. Following the derivation in~\cite{Basar:2014jua}, the modular expression of the twisted partition function can be expressed as
\begin{equation}
\tilde Z(\tau)=\eta^3(2\tau)  \prod _{i=1}^2 \left[ \frac{2\cos (\pi \nu_i)}{\vartheta_2(\nu_i| \tau)}    \right]. \label{eq.tZ2-tau}
\end{equation}
Accordingly the coefficient of the logarithmic term in the high-temperature expansion changes from $-3/2$ to $-1/2$.
When $n_f=4$ and $n_s=6$, $Q=1$ becomes a quartic root. Note such a special case is realized in ${\mathcal N}=4$ SYM. Setting the remaining two zeros as $Q_1=1/Q_4=-r$, the expression further simplifies
\begin{equation}
\tilde Z(\tau)=\frac{\eta^3(2\tau)}{\eta^3(\tau)} \left[ \frac{2\cos (\pi \nu)}{\vartheta_2(\nu| \tau)}\right].\label{eq.tZ3-tau}
\end{equation}
In the high-temperature expansion the coefficient of the logarithmic term now becomes positive, $1/2$. The corresponding thermal partition functions could be directly obtained through the transformation $\tau\to \tau+1/2$.

Interestingly, these special cases serve as the borderline for the different phase structure in the twisted theory. This is manifest from the fact that when $P(Q)=0$ has a root $Q\in(0,1)$, the Hagedorn instability appears. It is not difficult to find that the condition for the existence of such a root is
\begin{equation}
n_f<\frac{n_s}{2}+1,\quad n_f=\frac{n_s}{2}+1>4.\nonumber
\end{equation}
Correspondingly, the theory experiences a Hagedorn transition. Otherwise, the theory always stays in the confining phase, e.g., ${\mathcal N}=1$ SYM~\cite{Unsal:2007fb} and the special case $n_f\ne0, n_s=0$~\cite{Basar:2014jua}. In the thermal theory, the equation $P(-Q)=0$ always has a root $Q\in(0,1)$. As a result, a Hagedorn transition is unavoidable.

\section{Small-$|\tau|$ expansion in pure Yang-Mills theory}
\label{App:YM}
 Here we give the explicit derivation of the expansion (\ref{eq.ZYM-exp}) of pure Yang-Mills theory. The polynomial $P(Q)$ is now
\begin{equation}
P(Q)=Q^6-3Q^4-3Q^2+1.
\end{equation}
Since no fermions are introduced, we have $P(-Q)=P(Q)$. The roots are in reciprocal pairs due to the symmetry $Q\to 1/Q$ of the theory, and can be specified as
\begin{equation}
r_1=\mathi,\quad r_2=\sqrt{2+\sqrt{3}}, \quad r_3=-\sqrt{2+\sqrt{3}}.
\end{equation}
The corresponding values for $\nu_i$ are
\begin{equation}
\nu_1=\frac{1}{4},\quad \nu_2=\frac{\mathi \theta}{\pi},\quad \nu_3=\frac{1}{2}+\frac{\mathi \theta}{\pi},
\end{equation}
with $\theta= -\log (2+\sqrt{3})/4$ as in (\ref{eq.ZYM-exp}). The partition function can be expressed with these parameters in the twisted form as
\begin{equation}
\tilde Z_{YM}(\tau)=\eta^3(2\tau)  \prod _{i=1}^3 \left[ \frac{2\cos (\pi \nu_i)~\eta(\tau)}{\vartheta_2(\nu_i| \tau)}    \right],\label{eq.tZYM-tau}
\end{equation}
or in the thermal form
\begin{equation}
Z_{YM}(\tau)=\eta^3(2\tau+1) \prod _{i=1}^3 \left[ \frac{2\cos (\pi \nu_i)~ \eta(\tau+\frac{1}{2})}{\vartheta_2(\nu_i| \tau+\frac{1}{2})}    \right].\label{eq.ZYM-tau}
\end{equation}
The equivalence of the above two expressions may be slightly manifest if we combine the contributions from $r_2$ and $r_3$, which results
\begin{equation}
\tilde Z_{YM}(\tau)=\eta^4(2\tau) \eta(\tau)   \frac{2\cos (\pi \nu_0)}{\vartheta_2(\nu_0| 2\tau)} \frac{2\cos (\pi \nu_1)}{\vartheta_2(\nu_1| \tau)}   ,\label{eq.tZYM-tau2}
\end{equation}
 with $\nu_0=\frac{1}{2}+\frac{2\mathi \theta}{\pi}$. The small-$|\tau|$ expansion is easy to derive from (\ref{eq.tZYM-tau}) and (\ref{eq.tZYM-tau2}), by employing the following identities
 \begin{eqnarray}
 \eta(-\frac{1}{\tau})&=&\sqrt{-\mathi \tau} ~\eta(\tau),\nonumber\\
 \vartheta_4\left(\frac{\nu}{\tau}\Big|-\frac{1}{\tau}\right)&=&\sqrt{-\mathi \tau}~\mathe^{\mathi \pi \nu^2/\tau}~\vartheta_2(\nu|\tau).\nonumber
\end{eqnarray}
Here the fourth Jacobi's $\vartheta$ function is defined as
\begin{equation}
\vartheta_4(\nu|\tau)=\prod _{n=1}^\infty (1-\xi^{2n})(1-2~\cos (2\pi \nu)~\xi ^{2n-1}+\xi^{4n-2}),
\end{equation}
with $\xi\equiv \mathe^{\mathi \pi \tau}$. The derivation from (\ref{eq.ZYM-tau}) needs a little more effort. One will need the Landen transformation~\cite{Wang1989}
\begin{equation}
\vartheta_2^2(\nu|\tau)=\vartheta_2(2\nu|2\tau) \vartheta_3(0|2\tau) +\vartheta_2(0|2\tau) \vartheta_3(2\nu|2\tau),
\end{equation}
with the third Jacobi's $\vartheta$ function
\begin{equation}
\vartheta_3(\nu|\tau)=\prod _{n=1}^\infty (1-\xi^{2n})(1+2~\cos (2\pi \nu)~\xi ^{2n-1}+\xi^{4n-2}).
\end{equation}
Also the following identities are required
\begin{eqnarray}
\vartheta_2(\nu|\tau+1)&=&\mathe^{\mathi \pi/4} \vartheta_2(\nu|\tau),~~~\vartheta_3(\nu|\tau+1)=\vartheta_4(\nu|\tau),\nonumber\\
 \vartheta_2\left(\frac{\nu}{\tau}\Big|-\frac{1}{\tau}\right)&=&\sqrt{-\mathi \tau}~\mathe^{\mathi \pi \nu^2/\tau}~\vartheta_4(\nu|\tau).\nonumber
\end{eqnarray}
All the different expressions give the same expression (\ref{eq.ZYM-exp}).

\end{document}